\documentclass[prd,showpacs]{revtex4}

\usepackage{epsfig}
\usepackage{graphicx}
\usepackage{latexsym}
\usepackage{amsmath}

\newcommand\beq{\begin{equation}}
\newcommand\eeq{\end{equation}}
\newcommand\bea{\begin{eqnarray}}
\newcommand\eea{\end{eqnarray}}
\newcommand\FF{{\cal F}}

\newcommand\si{{\sigma}}
\newcommand\gtwid{\mathrel{
\raise.3ex\hbox{$>$\kern-.75em\lower1ex\hbox{$\sim$}}}}

\newcommand\bfA{{\mathbf A}}

\newcommand\bfs{{\mathbf s}}
\newcommand\bfx{{\mathbf x}}
\newcommand\bfeta{{\mbox{\boldmath $\eta$}}}

\begin{document}

\title{Interaction between vortices in models with two order parameters}

\author{R. MacKenzie}
\email{richard.mackenzie@umontreal.ca}
\author{M.-A. Vachon}
\email{vachon@lps.umontreal.ca}
\author{U. F. Wichoski}
\email{wichoski@lps.umontreal.ca}

\affiliation{Laboratoire Ren\'e-J.-A.-L\'evesque, Universit\'e de Montr\'eal\\
C.P. 6128, Succ. Centre-ville, Montr\'eal, QC H3C 3J7}

\begin{abstract}
The interaction energy and force between widely separated 
strings is analyzed in a field theory having applications to superconducting
cosmic strings, the SO(5) model of high-temperature superconductivity,
and solitons in nonlinear optics. The field theory has two order
parameters, one of which is broken in the vacuum (giving rise to
strings), the other of which is unbroken in the vacuum but which could
nonetheless be broken in the core of the string. If this does occur,
there is an effect on the energetics of widely separated strings. This
effect is important if the length scale of this second order parameter
is longer than that of the other fields in the problem.
\end{abstract}

\pacs{74.20.De,74.20.Mn,74.72.-h,11.27.+d,98.80.Cq,42.65.Tg}

\preprint{UdeM-GPP-TH-02-106}

\maketitle

\section{Introduction}

In at least three very different contexts, nonlinear effects in field
theories with spontaneous symmetry breaking
give rise to topological solitons wherein a second order
parameter (unbroken in the vacuum) attains a nonzero expectation value
in the core of the soliton. 

First, bosonic superconducting cosmic strings
\cite{Witten:1985eb} arise in a model with two $U(1)$ symmetries. The
first of these (which could be
gauged or ungauged) is spontaneously broken by a complex scalar
field $\varphi$; this gives rise to the possibility of string
solutions, where the phase of $\varphi$ changes by $2\pi$ around a
large loop in space. In the core of such a string, $\varphi\to0$.
The second symmetry is gauged and unbroken, and is identified with
electromagnetism. It is supposed that an electromagnetically charged
scalar field $\si$ exists; although its VEV is zero,
the potential $V(\si,\varphi)$ is such that if one
forces $\varphi$ to zero, then $V(\si,0)$ is minimized for
$\si\neq0$. Thus, it is possible that $\si$ attains an expectation
value inside the core of a string, making the string a superconducting
wire. Whether or not this actually occurs
is a detailed dynamical question.
\nocite{Lazarides:1985my,Copeland:1987th,
MacKenzie:1987ye,MacKenzie:1988yf,Hindmarsh:1988vh,Hill:1988qx,
Haws:1988ax,Davis:1989ij,Frieman:1990zj}
A variety of generalizations of this
idea have been discussed in
Refs.~\cite{Lazarides:1985my}-\cite{Frieman:1990zj}.

Second, in nonlinear optics, one can consider beams
for which the amplitude of the electromagnetic field envelope function
vanishes and its phase changes by an integer multiple of $2\pi$ at a
given point. In a nonlinear medium such a configuration can be a
stable solution of the equations of motion, known as an optical vortex
soliton (see \cite{Kivshar:1998} for references and review).
By coupling a second propagating mode to the first, the vortex can act
as a waveguide for the second mode, which is confined to the core of
the vortex (for theoretical work, see
\cite{Alexander:2000,Carlsson:2000}; for recent experimental results
see  \cite{Truscott:1999,Morin:1995,Kang:1995}).
The first and second modes are analogous to the fields $\varphi$ and
$\si$ above, respectively, although in the optical context both fields
are ungauged. (It should also be mentioned that the fields are not
really order parameters here.) 

Finally, the SO(5) model of high-temperature superconductivity (HTSC)
\cite{Zhang:1996} is a ``unification'' of the two phenomena that occur
in these materials at low temperature: antiferromagnetism (AF) at low
doping (including the undoped case) and superconductivity (SC) at
higher doping. The Ginzburg-Landau (GL) model is written in terms of a
five-component real order parameter composed of
a three-component field $\bfeta$
describing antiferromagnetism and a
two-component field $\phi$ describing superconductivity. In the SC
phase, the latter field attains an expectation value, and strings
(vortices) exist, just as in conventional (SO(2))
superconductivity. It is possible that $\bfeta$ attains a nonzero
expectation value inside the core, making the core AF,
which could provide an experimental test of the SO(5) model
\cite{Arovas:1997}.

In this paper, we will study a field theory which applies (with minor
modifications) to
any of the above situations, and will study
the energetics of two (widely-separated) vortices. In the cosmic
string context
this has an influence on the dynamics of a network of strings
\cite{Bettencourt:1994kc,Bettencourt:1995kf,Bettencourt:1997qe};
in optics this affects the stability of solitonic
waveguides \cite{Kivshar:1998}; and in the superconducting case this is 
one way to study the type of superconductor described by a given model, 
which provides another test of the SO(5) model
\cite{Juneau:2001vz,Juneau:2002ax}.

In the next section, we will establish notation and
review previous work (expressed in the language of
superconductivity, but easily translated into the other contexts),
wherein the effect of the AF phase on the magnetic behaviour was
studied, and the conditions under which SO(5) superconductors
are type I or II were determined. This was done in two ways. First,
the free energy $F$ of a
vortex as a function of its winding number $m$ was calculated
numerically \cite{Juneau:2001vz}.
If the energy per unit winding number
$\FF(m)\equiv F(m)/m$ is a decreasing
function of $m$, type I superconductivity results, since it
is then energetically preferable for a given flux to penetrate a
superconductor in one large region, inside which superconductivity is
destroyed. On the other hand, if $\FF(m)$ is
an increasing function of $m$, the superconductor is type
II, since the energetics then prefers a given amount of flux
to be divided into a network of
vortices of winding number $m = 1$. Second,
the type of superconductor
was deduced from a determination \cite{Juneau:2002ax} of
the critical magnetic fields (see, for example,
\cite{Tinkham:1996}). While this approach is perhaps less intuitive
than the previous one, it has the advantage that a simple analytical
determination of the
boundary between type I and type II superconductivity can be made.

A third way of differentiating between type I and type II
superconductors is through the force between two widely-separated
vortices: they
attract or repel for superconductors of type I or II,
respectively. This approach is very intuitive, and is also applicable
to both the cosmic strings and nonlinear optics contexts
(where the notion of critical magnetic fields which
restore the symmetry broken in bulk doesn't apply).
In this article, we will examine this third approach, adopting for the
most part the language of SO(5) superconductivity, though much of the
discussion can easily be exported to other contexts such as those
mentioned above.

The method used is essentially that used by Speight
\cite{Speight:1997sh}. We write an expression for the
energy of two vortices using a point vortex approximation, wherein the
full nonlinear free energy is replaced by a linearization of it plus
point sources for each vortex. The approximation is expected to be
valid if the intervortex separation is much larger than the core size.
The approximate free energy can be
written as the sum of the individual vortex energies and an
interaction energy, from which the intervortex force can easily be
found. In conventional
SC, vortices attract or repel one
another in type I and II SC, respectively. As we will see, this is
not always the case here: for certain values of the parameters of the
SO(5) model, a superconductor which is type I (in the thermodynamic
sense) will have repulsive, not attractive, vortices. This unusual
behaviour is unlikely to be seen in HTSC,
however, as it occurs only for small values of the GL parameter
($\kappa\sim o(1)$), whereas in all known HTSCs, $\kappa\gtwid50$.

\section{Review of previous work}

The model we wish to consider is described by the following
two-dimensional free energy:
\bea
\label{free}
\hat{F} &=& \int d^2x \left\{
\frac{(\nabla\times\hat{\bfA})^2}{8 \pi} 
+ \frac{ \hbar^2}{2 m^*}\left| \left( -i\nabla 
- \frac{e^*}{\hbar c}\hat{\bfA} \right) \hat\phi \right|^2\right.
\nonumber\\
&&\qquad\qquad\qquad\left.+ \frac{\hbar^2}{2m^*}( \nabla \hat\bfeta)^2 
+ V( \hat\phi, \hat\bfeta) \right\}.
\eea
Here $\hat\phi$ and $\hat\bfeta$ are the SC and
AF order parameters, respectively. The former is a complex field
associated with the U(1) gauge field $\hat{\bfA}$, while the
latter is a real triplet whose SO(3) symmetry is
ungauged.\footnote{Hats denote dimensionful quantities. The constants
  appearing in (\ref{free}) are appropriate to superconductivity;
  readers unfamiliar with these conventions will be relieved to learn
  that the notation will be streamlined presently, when we go to
  dimensionless variables.}

The potential is taken to be an even, quartic function
of $|\hat\phi|$ and $|\hat\bfeta|$:  
\begin{displaymath}
 V(\hat\phi,\hat\bfeta)= -\frac{{a_1}^2}{2} |{\hat\phi}|^2 -
 \frac{{a_2}^2}{2} |{\hat\bfeta}|^2 
+ \frac{b}{4}\left(|{\hat\phi}|^4 + 2|{\hat\phi}|^2 |{\hat\bfeta}|^2 
+ |{\hat\bfeta}|^4 \right).
\end{displaymath}
For simplicity, we have imposed SO(5) symmetry on the quartic
couplings; the quadratic couplings are negative to give rise to symmetry
breaking. The ground state depends on the value of the parameter
$\beta\equiv a_2^2/a_1^2$, and is SC if $\beta<1$. We are
primarily interested in this case; the ground state can be written 
$(|\hat\phi|,|\hat\bfeta|)=(v,0)$, where $v\equiv a_1/\sqrt{b}$.

The model can be simplified somewhat by rescaling the fields and the
position variable. Defining dimensionless (unhatted) quantities
\[
\hat{\bfA}={a_1 c\sqrt{m^*}\over e^*}\bfA,\quad
\hat\phi=v\phi,\quad
\hat\bfeta=v\bfeta,\quad
\bfx=\sqrt{{m^*c^2\over4\pi{e^{*}}^2v^2}} \bfs,\quad
\hat F={a_1^2c^2m^*\over4\pi {e^{*}}^2} F,
\]
we have the dimensionless free energy
\bea
\label{free2}
F&=&\frac12\int d^2s\left\{(\nabla\times \bfA)^2
+{1\over\kappa^2}\left(\left|(-i\nabla-\kappa \bfA)\phi\right|^2
+(\nabla\bfeta)^2\right)\nonumber\right.\\
&&\qquad\qquad\qquad
\left.-\phi^2-\beta\eta^2+\frac12(\phi^2+\eta^2)^2 +{1 \over 2}\right\},
\eea
where $\phi=|\phi|$, $\eta=|\bfeta|$, the derivatives are now with
respect to $\bfs$,
and
\[
\kappa=\sqrt{{b\over4\pi}}{m^*c\over\hbar e^*}
\]
is the usual GL parameter. (A constant has been added to $F$ so that
the ground state has zero energy.)

From (\ref{free2}), we see that the behaviour of
the model is completely determined by
two dimensionless parameters: $\kappa$ and the parameter $\beta$,
which is the ratio of the quadratic coefficients of $\eta$ and $\phi$.

An observation which will be useful below is that if we set $\eta=0$,
then the free energy (a function of $\phi$ and $\bfA$ only) reduces to
that of the SO(2) model. Thus, for example,
the energy of a static configuration for which $\eta=0$ in the SO(5)
model will be exactly equal to the energy of the same field
configuration in the SO(2) model with the same value of
$\kappa$.

Since the complex field $\phi$ attains a nonzero VEV, vortex solutions
exist, wherein the phase of $\phi$ changes by $2\pi$ around a large
circle in space. One can also consider configurations of higher
winding number; a rotationally symmetric ansatz of winding number $m$
(``$m$-vortex'')
is essentially the conventional one for
$\phi$ and $\bfA$ together with a rotationally invariant
ansatz for $\bfeta$, whose
orientation is taken to be a fixed,
arbitrary unit vector $\hat{\mathbf{e}}$:
\beq
\phi(\bfs)=\varphi(s) e^{im\theta},
\quad {A_i}(\bfs)=\epsilon_{ij} \frac{s_j}{s} A(s),
\quad \bfeta(\bfs) = \hat{\mathbf{e}}\eta(s). 
\label{ansatz}
\eeq
As in the conventional case, these solutions carry a magnetic flux
proportional to $m$ (specifically, the dimensionless flux is
$\Phi=-2\pi m/\kappa$).

As mentioned in the Introduction, a non-vanishing $\eta$ 
field may appear inside the core of the vortex due to the competition 
between the potential and gradient energy terms. By examining the
potential, one can see qualitatively that for fixed $\kappa$,
as $\beta$ increases, the
impetus for $\eta$ to be nonzero in the core of a vortex grows, since
the potential energy savings gained by having $\eta\ne0$ increases
with
$\beta$, while the kinetic energy cost is independent of $\beta$.
Thus, we can define a critical value of $\beta$, above which the core
of a vortex is AF and below which it is normal.
This critical value depends upon $\kappa$, and also
upon the winding number $m$ of the vortex. (The dependence on $m$ can
also be argued qualitatively, by noting that the larger the winding
number, the greater the region in which $\varphi$ is nearly zero, and
the greater the impetus for $\eta$ to be nonzero.) $\beta_{AF}(\kappa,m)$
was found numerically for $m=1$ to 5 in Ref. \cite{Juneau:2001vz}.

The fact that the model reduces to the SO(2) model if $\eta=0$ enables
us to easily make contact with that model; we need only set
$\beta$ to zero, since then there is never any reason for $\eta$ to be
nonzero in all situations we will consider here.

Substituting the ansatz (\ref{ansatz}) into (\ref{free2}),
we find the following free energy:
\begin{eqnarray}
\label{fs1}
F(m)&=&\pi\int s\,ds \Biggl\{ 
\left( A' + \frac{A}{s} \right)^2
+ \frac{1}{\kappa^2} \left[ \varphi'^2 + 
 \left(\frac{m}{s} + \kappa A \right)^2 \varphi^2
+ \eta'^2 \right] 
\nonumber\\
& &\qquad\qquad\qquad -\varphi^2 - \beta \eta^2 
+ \frac{1}{2}( \varphi^2 + \eta^2)^2 + \frac{1}{2} \Biggr\}.
\end{eqnarray}
The free energy, written as a function of $m$, depends also, of
course,
on the two parameters of the model,
$\kappa$ and $\beta$.

The equations of motion that follow from (\ref{fs1}) are
\begin{equation}\label{eq1}
 \frac{1}{\kappa^2} \left[ \varphi'' + \frac{1}{s}\varphi' 
- \left(\frac{m}{s} +\kappa A
\right)^2\varphi \right] +\varphi(1-\varphi^2 -\eta^2)=0,
\end{equation}
\begin{equation}\label{eq2}
\frac{1}{\kappa^2}\left( \eta'' + \frac{1}{s}\eta'\right) 
+ \eta(\beta - \varphi^2 -\eta^2)=0,
\end{equation}
\begin{equation}\label{eq3}
A'' + \frac{1}{s}A' - \frac{1}{s^2}A 
- \left( \frac{m}{ks} + A\right)\varphi^2=0.
\end{equation}

These equations cannot be solved analytically; however, asymptotic
solutions for large $s$ can be found. Defining
\beq\label{fas}
\varphi(s)=1-f(s)\quad
\mbox{and}\quad
A(s)=-{m\over\kappa s}+a(s),
\eeq
the fields $f$, $a$ and $\eta$ approach zero exponentially
as $s\to\infty$.
The linearized equations in these fields have solutions
\begin{equation}\label{asym}
f(s) = C_f K_0(\sqrt{2}\kappa s), \quad 
a(s)= C_a K_1(s), \quad 
\eta(s) = C_\eta K_0(\sqrt{1-\beta}\kappa s)
\end{equation}
where $K_0$ and $K_1$ are modified Bessel functions of the second kind
and the $C$'s are constants not determined by the linear equations.
These results 
will be useful for deriving the potential energy between a pair of
widely-separated
vortices in the next section.
 
By studying either the energetics of vortices as a function of their
winding number \cite{Juneau:2001vz} or by studying the critical magnetic 
fields \cite{Juneau:2002ax}, one can determine a curve
in the $\beta$-$\kappa$ plane indicating the boundary between type I and
type II behaviour, which yields a surprising result.

To see this, let
us first recall the situation in the conventional SO(2) model,
focusing our attention on vortex energetics.
There, the only parameter is $\kappa$, and
$\kappa=1/\sqrt2\equiv\kappa_c$ is the critical value separating type I
($\kappa<\kappa_c$) and type II ($\kappa>\kappa_c$) behaviour. For
example, one finds that
the energy per unit magnetic flux of a vortex
$\FF(m)=F(m)/m$ decreases or increases
with $m$ according to whether $\kappa<\kappa_c$ or
$\kappa>\kappa_c$. In either case, $\FF(m)$ is a monotonic function of
$m$, tending towards a constant value for large $m$.

In the SO(5) case, the monotonicity of $\FF(m)$ is no longer
guaranteed. The reason is that for fixed $\kappa$ and $\beta$, as $m$
increases one can go from a normal vortex core to an AF core. As long
as the core is normal, its energy is exactly as in the SO(2) model, as
can be seen from (\ref{free2}). However, as $m$ increases, eventually
the core becomes AF, at which point the energetic picture changes: in
particular, the energy of all subsequent $m$-vortices will be reduced
(relative to the equivalent SO(2) model),
lessening the degree to which the superconductor is type II, and in
some cases even changing the superconductor from type II to type
I. For fixed $\kappa$, this effect is stronger the larger the value of
$\beta$; we can therefore
define a critical value $\beta_c(\kappa)$ as that value for which
$\FF(m)$ goes from an increasing (type II) to a decreasing
(type I) function of $m$, in the limit of large $m$.

It is strange, however, that (in contrast to conventional SC) $\FF(m)$
is not necessarily a monotonic function: it could increase with $m$
while the core is normal and decrease with $m$ subsequently. The large-$m$
behaviour indicates a type I superconductor, yet the energy of a
2-vortex is more than double that of a single vortex, which is a
feature of type II superconductivity. This does not cast doubt on the
fact that such a SC is indeed type I, in a thermodynamic sense;
however, it will certainly affect vortex dynamics.

\begin{figure} 
\begin{center}
\includegraphics[width=5cm, angle=-90]{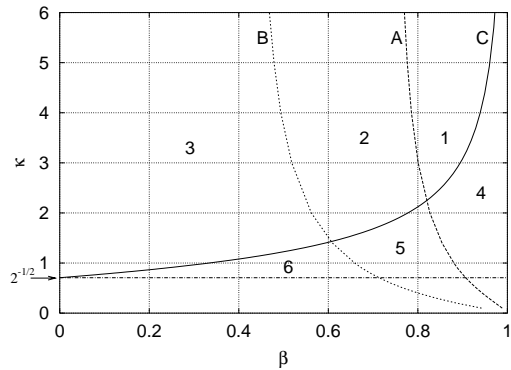}
\end{center}
\caption{\label{kbplane}Stability of vortex lattice in the SO(5)
  model.
Curves A and B 
separate the normal core region (left) from the AF core region (right)
in the case of 
$m=1$ and $m=2$ vortices, respectively. Curve C depicts the border between 
type I and type II behaviour
(see (\ref{kc})). These curves divide the $\beta$-$\kappa$ 
plane into sevelar regions, the most interesting of which are regions
  5 and 6, where the SC is type I yet vortices (of winding number 1)
  repel one another.}
\end{figure}

The behaviour at large $m$
is confirmed by studying the critical magnetic fields
$H_c$ and $H_{c2}$. As in conventional SC, the relative magnitude of
these is an indicator of the type of SC. In \cite{Juneau:2002ax} the
critical fields were found to be given by the following expressions:
\begin{equation}
H_c(\beta)= H_c^0 \sqrt{1- \beta^2}, \qquad H_{c2}(\beta) = \sqrt{2} 
\kappa H_c^0 ( 1- \beta),
\end{equation}
where $H_c^0$ is the thermodynamic critical field in the conventional SC 
model. The border between type I and type II will then be given by equating 
these two fields, leading to the following expressions for
$\beta_c(\kappa)$ and, by inversion, $\kappa_c(\beta)$ (that value of
$\kappa$ above which a SC is type II, for fixed $\beta$):
\begin{equation}\label{kc}
\beta_c(\kappa)={2\kappa^2-1\over2\kappa^2+1},
\qquad
\kappa_c (\beta) = (1/ \sqrt{2})\sqrt{(1 + \beta)/( 1- \beta)} \, .
\end{equation}
This line, as well as the lines separating the normal/AF core boundary
for $m=1$ and $m=2$ vortices, lines C, A, and B respectively, are shown 
in Figure \ref{kbplane}.

\section{Intervortex force}

As mentioned in the Introduction, a third way of determining the
magnetic nature of a superconductor is through the interaction
energy and force between vortices. We will
study the interaction of widely-separated
vortices in the SO(5) model. In conventional superconductivity,
vortices of any separation are attractive or repulsive in the case of
type I or type II superconductors, respectively, as was shown in great
detail numerically by Jacobs and Rebbi \cite{Jacobs:1979ch}. We would
thus expect the intervortex force in the SO(5) model
to change from repulsive to
attractive as we cross curve C in Figure \ref{kbplane}
from left to right. This turns out to be not always the case. The
reason is that vortex energetics does not always reflect the type of
superconductor in the SO(5) model, as a  comparison of the
following two facts demonstrates. On the one hand,
the type of an SO(5) superconductor depends critically on the AF
sector (see \cite{Juneau:2002ax}). On the other hand,
when vortices have normal
cores, their behaviour (their interaction energy, in particular) is
blind to the AF sector of the model, as was noted in the previous
section. Thus, vortices with normal cores cannot be expected to
necessarily behave in the fashion dictated by the type
of superconductor involved. In particular, it is possible that
vortices in a type I (in the sense of bulk
thermodynamic properties) superconductor repel one another.

One subtlety which arises in the interaction of vortices once their 
cores become AF is that while the orientation of $\boldsymbol{\eta}$ 
can be taken fixed in the core of any given vortex, in the absence of 
anisotropies, there is no reason to expect the AF cores of different 
vortices to be oriented in the same direction. When it becomes 
energetically favourable to develop an AF core, each vortex randomly 
selects a drection in which $\boldsymbol{\eta}$ will point. The 
interaction energy between vortices will depend on the angle between 
the two AF order parameters. As we will see, the contribution of 
$\boldsymbol{\eta}$ to the vortex interaction energy is most important 
if the two orders parameters are parallel. Our discussion will assume 
for the most part that this is the case. 

To clarify the situation, consider the interaction of ordinary
vortices (that is, vortices of winding number 1). The core of these
vortices is normal or AF to the left or right of curve A of Figure
\ref{kbplane}, respectively. Their interaction energy will be shown
to obey the following behaviour. If the core is normal (to the left of
curve A), the energetics is exactly as dictated by the SO(2) model:
attractive for $\kappa<1/\sqrt2$ and repulsive for
$\kappa>1/\sqrt2$. If the core is AF (to the right of curve A), then
(except for a very thin transition zone wherein the AF core turns on) the
energetics correctly reflects the type of superconductor involved
(repulsive or attractive to the left or right of curve
C). 

If we imagine fixing $\kappa$ and increasing $\beta$, then there
are three cases. First, if
$\kappa<1/\sqrt2$ nothing dramatic happens: for all $\beta$ the model
describes type I superconductivity
and vortices attract. Eventually curve A is crossed, so the
vortices develop AF cores; they remain attractive, however.

Second, if $\kappa$ lies above its
value at the point of intersection of curves A and C
($\kappa>2.25$), the situation is slightly more interesting: the core 
turns AF to the right of curve A, but
the vortices remain repulsive until curve C is crossed, after which
the superconductor is type I and the vortices attract.

The third,
intermediate case, where curve C is crossed before curve A as $\beta$
increases ($1/\sqrt2<\kappa<2.25$), is the
most interesting. When curve C is crossed, the superconductor becomes
type I, so the vortices {\em ought to} attract. However, their cores
are normal, their energetics is as in the SO(2) model, and they repel,
since $\kappa>1/\sqrt2$. This anomalous behaviour persists until curve
A is crossed, when the core becomes AF and the vortices start to
attract one another -- the expected behaviour for a type I
superconductor. 

In summary, curve C delineates the boundary between type I (to the 
right) and type II (to the left) superconductivity; however, 
in regions 5 and 6 the superconductor 
has a sort of identity crisis: it is type I, but its vortices behave 
as type II vortices.\footnote{Perhaps some clarification on what is 
meant by type I vs type II superconductivity is warranted. In 
conventional superconductivity, type II superconductivity (for 
instance) is characterized by $\kappa>1/\sqrt2$, negative surface 
energy at a normal/superconducting boundary, repulsive vortices, and 
$H_{c2}>H_c$. These are so inextricably connected that any of these 
features could be used as a definition of type II; the rest would 
follow. As we have seen, and as will be argued below (see also 
\cite{Juneau:2002ax}), these features no longer imply one another in 
SO(5) superconductivity. We refer to a superconductor as type I or 
II according to whether the surface energy is negative or positive, or 
equivalently, according to the relative value of $H_c$ and $H_{c2}$.} 

One can also study the interaction energy of vortices of higher 
winding number. The behaviour is qualitatively similar to that for 
1-vortices, although the region of disagreement between vortex 
energetics and the type of superconductor is reduced. For 2-vortices, 
for example, the core turns AF at a lower value of $\beta$ for any 
given $\kappa$ (curve B 
rather than A), and consequently it is only in region 6 that one finds
disagreement. This trend continues as the winding number is increased.

With this summary of the main results concluded, let us describe the
calculations involved in determining the interaction energy of a pair
of vortices. For arbitrary separation, an analytic solution is not
possible; even numerically, the problem is extremely difficult.
However,
the interaction between widely separated vortices is considerably more
tractable, and is clearly of interest, in that it is an important
ingredient in determining the
dynamics of widely separated vortices, and is normally directly
related to the type of superconductor involved.

If the vortices are widely separated, one can argue that the details
of the fields in the core ought to be unimportant, and we can model
them by a simplifying approximation, following
Speight \cite{Speight:1997sh}: we
can consider a linearized theory with point sources added at the
location of the vortices in such a way that
the long-range fields produced by the point sources in the linear
theory agree with those of the vortices in the original theory.

To derive the linearized theory, we must expand the free energy 
(\ref{free2}) around the asymptotic values of the fields. Before doing
this, however, it is useful to eliminate the phase of $\phi$, as this
degree of freedom becomes the longitudinal gauge field. Thus, we
take $\phi = 1 - f$ real,
and expand (\ref{free2}) in powers of $f$,
$\mathbf{A}$ and $\boldsymbol{\eta}$, up to quadratic 
terms. This results in:
\beq\label{hf}
F_{\rm free} ={1\over 2}\int d^2s \left\{
 \left( \nabla \times \mathbf{A}\right)^2+\mathbf{A}^2
 +{1\over\kappa^2}\left(\nabla f \right)^2+2f^2+{1\over\kappa^2}
\left(\nabla \boldsymbol{\eta} \right)^2 +
(1-\beta)\boldsymbol{\eta}^2 \right\}.
\eeq
To this, we must add couplings to sources:
\beq \label{hs}
F_{\rm source} = \int d^2s (\rho f +\boldsymbol{\sigma}\cdot 
\boldsymbol{\eta} + 
 \mathbf{j}\cdot\mathbf{A}).
\eeq
The equations of motion which follow from
$F_{\rm free}+F_{\rm source}$ are 
\beq\label{eq4}
\nabla^2 f - 2 \kappa^2f = \kappa^2 \rho,
\eeq
\beq\label{eq5}
\nabla^2\boldsymbol{\eta} - (1-\beta)\kappa^2 \boldsymbol{\eta} = 
\kappa^2 \boldsymbol{\sigma},
\eeq\beq\label{eq6}
\nabla^2 \mathbf{A} - \mathbf{A} = \mathbf{j}.
\eeq

The sources are to be chosen to give rise to fields which coincide
with the asymptotic vortex fields, after elimination of the phase of
$\phi$ via a gauge transformation. These asymptotic fields are (see
(\ref{asym})):
\beq
f(\bfs) = C_f K_0(\sqrt{2}\kappa s), \quad 
A_i(\bfs)= \epsilon_{ij}{s_j\over s}C_a K_1(s), \quad 
\boldsymbol{\eta}(\bfs) = \hat{\mathbf{e}} C_\eta K_0(\sqrt{1-\beta}\kappa s).
\label{asym2}
\eeq

We require sources such that the solutions of (\ref{eq4}-\ref{eq6}) 
are (\ref{asym2}). For $f$ and $\eta$ the answer is found directly by
substitution once we recognize that
\beq\label{poisson}
(\nabla^2 -\mu^2)K_0(\mu s) =-2 \pi \delta^2(\bfs);
\eeq
we find
\beq\label{cf}
\rho(\bfs) =- 2 \pi \frac{C_f}{\kappa^2} \delta^2(\bfs), \qquad
\boldsymbol{\sigma}(\bfs)= -2 \pi \hat{\mathbf{e}} 
\frac{C_{\eta}}{\kappa^2} \delta^2(\bfs). 
\eeq
For the gauge field, differentiation of (\ref{poisson}) and
substitution yields
\beq
j_i(\bfs)=2\pi C_a \epsilon_{ij}\partial_j\delta^2(\bfs).
\label{ca}
\eeq
The vortex is now described by point sources of magnitudes
$2 \pi C_f/\kappa^2$ and $2 \pi C_{\eta}/\kappa^2$ for $f$ and $\eta$,
respectively, and an infinitesimal current loop of magnetic dipole 
moment $ 2 \pi C_a$ for $\mathbf{A}$.

We are now in a position to derive the interaction energy of two widely 
separated vortices. Suppose the vortex positions are
$\mathbf{s}_1$ and $\mathbf{s}_2$. In the point vortex
approximation, each vortex is described by sources of the form
(\ref{cf},\ref{ca}), displaced to the position of the
vortex. Linearity of the equations of motion then implies that the
fields will just be the sum of the individual vortex fields, and the
energy of the configuration will be given by $F_{\rm free}+F_{\rm
  source}$, with fields $(f,\boldsymbol{\eta},\mathbf{A})
 = (f_1+f_2,\boldsymbol{\eta}_1 + \boldsymbol{\eta}_2,\mathbf{A}_1 + 
\mathbf{A}_2)$ and 
sources $(\rho,\boldsymbol{\sigma},\mathbf{j})=(\rho_1+\rho_2,
\boldsymbol{\sigma}_1 + \boldsymbol{\sigma}_2,\mathbf{j}_1+\mathbf{j}_2)$, 
where subscript 1,2 indicates $\bfs\to\bfs-\bfs_{1,2}$.
We can subtract off the
energy of each vortex to obtain the following interaction energy:
\beq
F_{\textrm{\footnotesize{int}}} = \int d^2s 
\left( \rho_1 f_2 + \boldsymbol{\sigma}_1 \cdot \boldsymbol{\eta}_2 + 
\mathbf{j}_1\cdot \mathbf{A}_2 \right)\equiv F_f + F_{\eta}+F_A.
\eeq
To evaluate $F_f$ we simply substitute $\rho_1=\rho(\bfs-\bfs_1)$ and
$f_2=f(\bfs-\bfs_2)$; we find
\beq
F_f = \int d^2s\left( -2 \pi \frac{C_f}{\kappa^2} 
\delta^2(\bfs-\bfs_1)\right) C_f K_0(\sqrt{2} \kappa|\bfs -\bfs_2|) 
= - 2 \pi \frac{C_f^2}{\kappa^2}K_0(\sqrt{2} \kappa d),
\eeq
where $d=|\bfs_1-\bfs_2|$ is the separation of the vortices.
The same argument applies 
to $F_{\eta}$, yielding
\beq
F_{\eta}= - 2 \pi \frac{C_\eta^2}{\kappa^2} \hat{\mathbf{e}}_1 
\cdot \hat{\mathbf{e}}_2 K_0(\sqrt{1-\beta}\kappa d).
\eeq
For $F_a$, a similar (but slightly more complicated) procedure leads to
\begin{eqnarray}
F_a & = & \int d^2 s \left(2 \pi C_a \epsilon_{ij}\partial_{s_j}
\delta^2(\bfs-\bfs_1)\right)
\epsilon_{ik}{(s-s_2)_k\over|\bfs-\bfs_2|}
C_a K_1(|\bfs-\bfs_2|) \nonumber \\
& = & 2 \pi C_a^2 {\partial\over\partial {s_1}_j}
\left({(s_1-s_2)_k\over|\bfs_1-\bfs_2|}
C_a K_1(|\bfs_1-\bfs_2|) \right) \nonumber\\
& = & 2 \pi C_a^2 K_0(d).
\end{eqnarray}

\begin{figure}[ht] 
\begin{center}
\includegraphics[width=5cm,angle=-90]{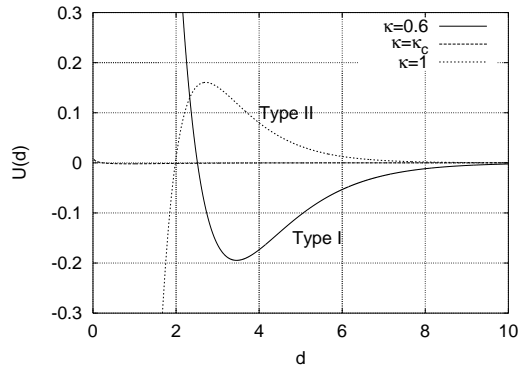}
\end{center}
\caption{\label{figso2}Behaviour of the potential in the case of 
the S0(2) model ($\beta = 0$) for various values of $\kappa$: 
$\kappa = 0.6$ (type I), $\kappa =\kappa_c$ (critical), and
$\kappa = 1.0$ (type II).
Note that as $d$ approaches zero, the exponential behaviour is 
lost and the curves can not be trusted anymore.}
\end{figure}

Thus, the interaction energy of two widely separated vortices 
takes the form
\beq\label{pot}
U(d) = 2\pi \left(C_a^2K_0(d)
 -\frac{C_f^2}{\kappa^2}K_0(\sqrt{2}\kappa d)
 -\frac{C_\eta^2}{\kappa^2} \hat{\mathbf{e}}_1 \cdot 
\hat{\mathbf{e}}_2 K_0(\sqrt{1-\beta}\kappa d) \right). 
\eeq

The first two terms give the interaction energy in the SO(2) case, and
coincide with the results of  \cite{Kramer:1971,Speight:1997sh}. The third 
term is the effect of the AF cores. As can be seen, it depends on the angle 
between the two order parameters. If they are parallel the contribution to the 
interaction energy is maximally negative, resulting in the greatest attraction, 
while antiparallel order parameters yield a positive (repulsive) force. If the 
order parameters are orthogonal the AF contribution to the interaction energy 
is zero and the results coincide with the SO(2) case. Since the parallel case 
is the most interesting and is also energetically preferred, we will 
assume in what follows that the order parameters are parallel, that is, that 
$\hat{\mathbf{e}}_1 \cdot \hat{\mathbf{e}}_2 = 1$.

In order to obtain useful information from (\ref{pot}), we must
determine the constants $C_a$, $C_f$ and $C_{\eta}$ by comparing
(\ref{asym}) with numerical solutions of the nonlinear equations
(\ref{eq1}-\ref{eq3}), which have been obtained previously
\cite{Juneau:2001vz,Juneau:2002ax}.

We will first study the conventional (SO(2)) case to make contact with
previous work; This
can be achieved within the SO(5) model by taking $\beta=0$.
Subsequently, we will examine
the general SO(5) case given by (\ref{pot}).
The coefficients $C_a$, $C_f$ are displayed for a variety of $\kappa$
(and for $\beta=0$) in Table 1.

\begin{table}[h]
\begin{center}
\begin{tabular}{|c|c|c||c|c|c|}
\hline
$\kappa$ & $C_a$ & $C_f/ \kappa$ & $\kappa$ & $C_a$ & $C_f/ \kappa$\\
\hline
\hline
0.4 & 6.561 & 2.888    &     1.0 & 1.417 & 2.325 \\
0.5 & 4.301 & 2.635    &     1.5 & 0.822 & 2.851 \\
0.6 & 3.126 & 2.486    &     2.0 & 0.534 & 5.061 \\

\bf{0.707} & \bf{2.388} & \bf{2.388} &  2.5 &0.402 & 12.972  \\

0.8 & 1.969 & 2.340    &     3.0 & 0.304 & 41.688 \\
0.9 & 1.651 & 2.318 &&& \\
\hline
\end{tabular}
\caption{Values of $C_a$ and $C_f/\kappa$ for various values of $\kappa$ in 
the case where the field $\eta = 0$ (achieved by setting $\beta$ to zero).}
\end{center}
\end{table}

There are three cases, depending on the value of $\kappa$.
The first case is $\kappa =\kappa_c$.
At this value, the potential energy is exactly zero for all separation. 
Indeed, the argument of both Bessel functions takes the same
value. Furthermore, the constants $C_a$ and $C_f/\kappa$ are equal,
as can be see numerically from 
Table 1 and also analytically from the following
argument. At $\kappa=\kappa_c$, we can write the free energy in a form
due to Bogomol'nyi \cite{Bogomol'nyi:1976}; the field equations can
then be written as first order differential equations:
\beq\label{bogol}
\varphi' - \left( \frac{m}{s} + \kappa A \right)\varphi = 0 \quad \mbox{and}
\quad \kappa\left(A' + \frac{A}{s} \right) + \frac{1}{2}( 1- \varphi^2)=0.
\eeq
Linearizing the first equation, we obtain $f'=-a$; substituting (\ref{asym}) 
into that equation, we find that the second Bogomol'nyi equation is satisfied 
if $\sqrt{2}C_f =C_a$. This argument is only valid at $\kappa_c$ because 
(\ref{bogol}) are not valid for other values of $\kappa$.
The potential is shown in Figure \ref{figso2}. 

The second case is $\kappa > \kappa_c$, corresponding to (conventional)
type II SC. In this case, we note that $K_0(\sqrt{2}\kappa d)$ falls off
more rapidly than $K_0(d)$ so the positive term in (\ref{pot}) will dominate 
over the negative one for large enough separation, no matter what the
values of the constants  
$C_a$ and $C_f/ \kappa$. Therefore, as is well known,
in conventional type II SC vortices are repulsive and give rise to stable 
vortex lattice. The resulting
potential is displayed in Figure \ref{figso2} for $\kappa= 1$.

The third case is $\kappa < \kappa_c$, corresponding to 
type I SC. Here, the situation is reversed:  $K_0(d)$
falls off more rapidly than $K_0(\sqrt{2}\kappa d)$, and we
conclude that the long-range potential is attractive. Thus, if we
start with an initial configuration formed by a number of
widely-separated vortices, ultimately they
will collapse into a single vortex of large winding number. 
This case is illustrated in Figure \ref{figso2} for $\kappa = 0.6$
The result (\ref{pot}) and the above discussion are in full agreement with 
\cite{Bettencourt:1995kf}, \cite{Speight:1997sh} and \cite{Jacobs:1979ch}.

Let us now turn our attention to the SO(5) case.
The coefficients $C_a$, $C_f$ and $C_\eta$
are displayed for a variety of $\kappa$
and $\beta$ in Table 2. 

\begin{table}[h]
\begin{center}
\begin{tabular}{|c|c|c|c|c||c|c|c|c|c|}
\hline
$\kappa$ & $\beta$ & $C_a$ & $C_f/ \kappa$ & $C_{\eta}/ \kappa$&
$\kappa$ & $\beta$ & $C_a$ & $C_f/ \kappa$ & $C_{\eta}/ \kappa$ \\
\hline
\hline
1.0 & 0.000 & 1.417 & 2.325 & 0.000 & 3.0 &0.00 & 0.362 & 41.688 & 0.000\\
    & 0.881 & 1.419 & 2.401 & 0.049 &     &0.81 & 0.362 & 45.742 & 0.083\\
    & 0.882 & 1.422 & 2.604 & 0.081 &     &0.82 & 0.364 & 52.005 & 0.119\\
    & 0.884 & 1.428 & 3.382 & 0.122 &     &0.83 & 0.364 & 61.320 & 0.145\\
    & 0.886 & 1.434 & 4.765 & 0.152 &     &0.85 & 0.368 & 74.346 & 0.185\\
    & 0.888 & 1.440 & 6.915 & 0.177 &     &0.87 & 0.371 & 88.687 & 0.216\\
    & 0.890 & 1.446 & 10.044& 0.198 &     &0.90 & 0.379 &132.886 & 0.249\\
    & 0.900 & 1.481 & 52.036& 0.281 &     &     &       &        & \\
\hline
\end{tabular}
\caption{Values of $C_a$, $C_f/\kappa$, and $C_{\eta}/\kappa$ for various 
values of $\kappa$ and $\beta$ for $m=1$ in the case of the SO(5) model.}
\end{center}
\end{table}

The possibility of an AF core gives rise to
the surprising results mentioned at the beginning of this section. 
As has already been noted,
if the core of the vortices is normal, there is no difference between
the energetics of the SO(5) model and the SO(2) model, since
the third term of (\ref{pot}) is zero.
Thus, the interaction energy of a pair SO(5) vortices (of winding
number 1) is unchanged from that described above, to the left of curve
A in Figure \ref{kbplane}.
To the right of this curve, however, the $\eta$ field is nonzero and
the third term makes an attractive contribution to the interaction
force, either lessening the degree to which the vortex energetics
behave in a type II manner, increasing the degree to which it behaves
in a type I manner, or (the most interesting possibility)
changing the vortices from repulsive to attractive.

We can read at which point this occurs from (\ref{pot}). The
force between widely separated vortices
depends on which of the fields is longest in range: the
gauge field, which provides a repulsive force, or one of the scalar
fields, whose forces are attractive. The range of the gauge field is
1, while that of the $\eta$ field (the longest range of the two scalar
fields, assuming it is nonzero in magnitude)
is $(\sqrt{1-\beta}\kappa)^{-1}$. These forces are equal for a certain
$\beta$, which we denote $\beta^*$:
\beq
\beta^*=1-1/\kappa^2.
\label{betastar}
\eeq
The long-range vortex
potential will be repulsive if $\beta<\beta^*$ and attractive if
$\beta>\beta^*$.

Note that
the caveat ``assuming it is nonzero in magnitude'' is essential: if the
range of $\eta$ is longer than that of $A$, but $C_\eta=0$, it will
obviously not have an effect on the vortex interaction.

\begin{figure}[t]
\begin{center}
\includegraphics[width=4cm,angle=-90]{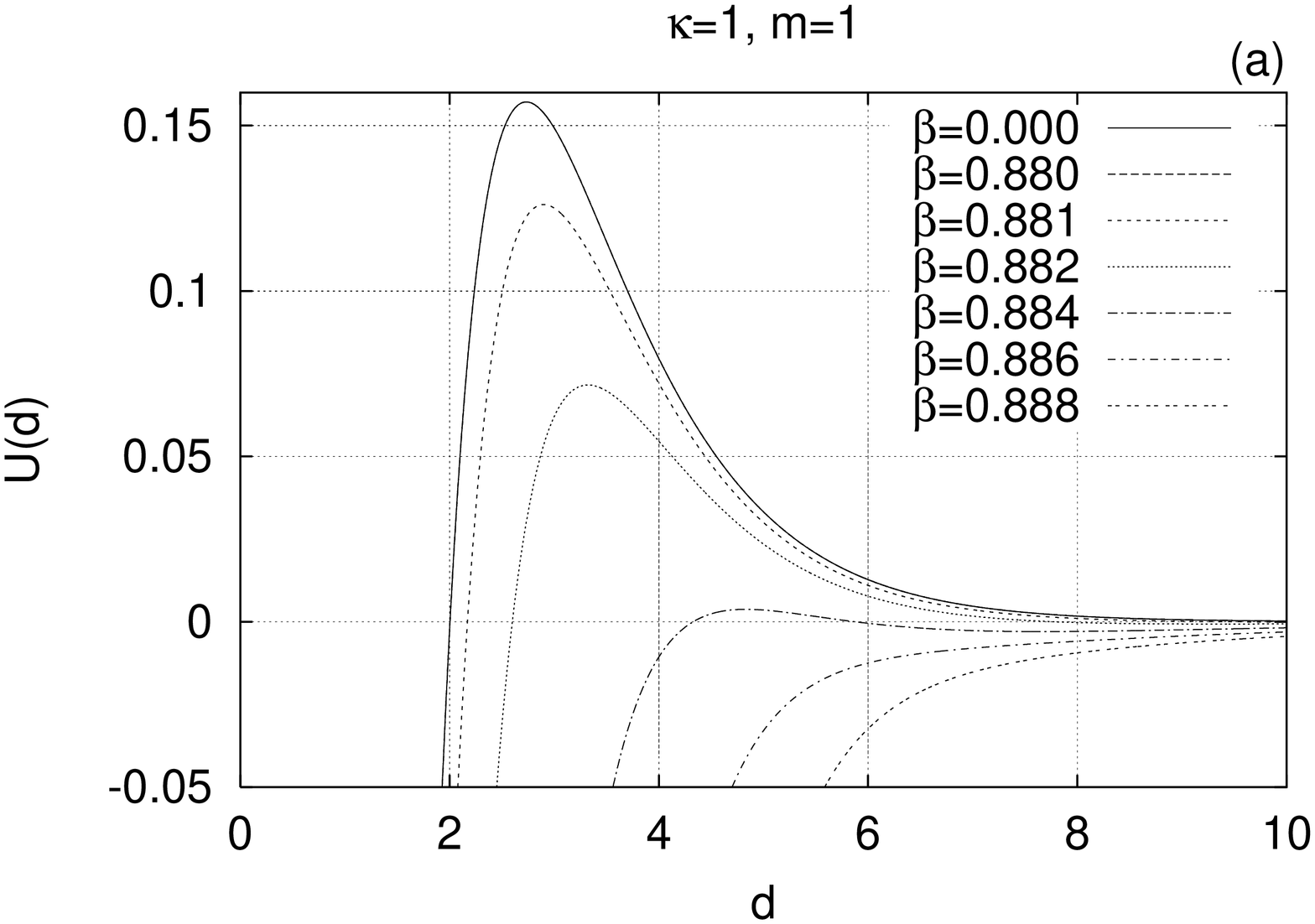}
\includegraphics[width=4cm,angle=-90]{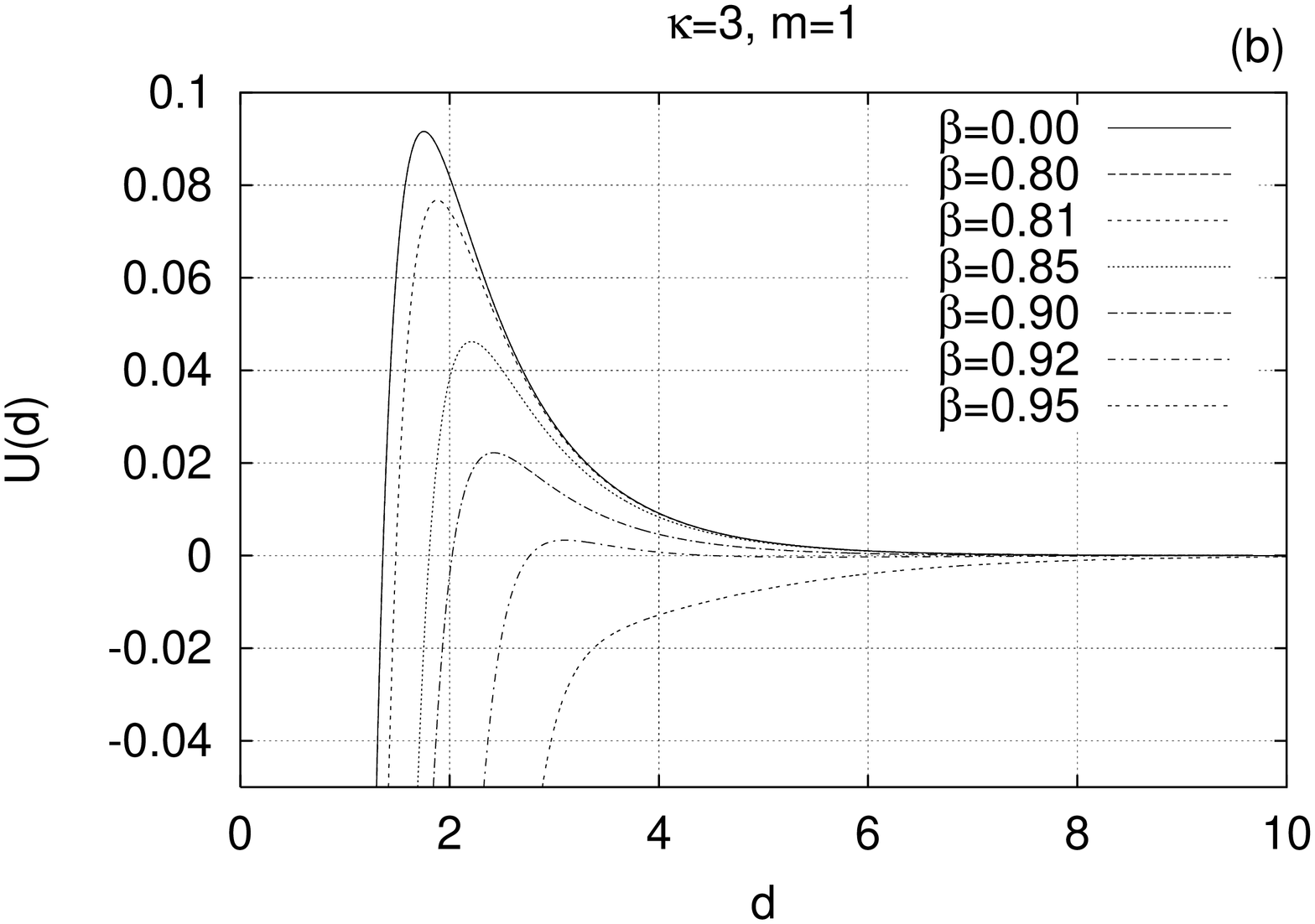} \\
\includegraphics[width=4cm,angle=-90]{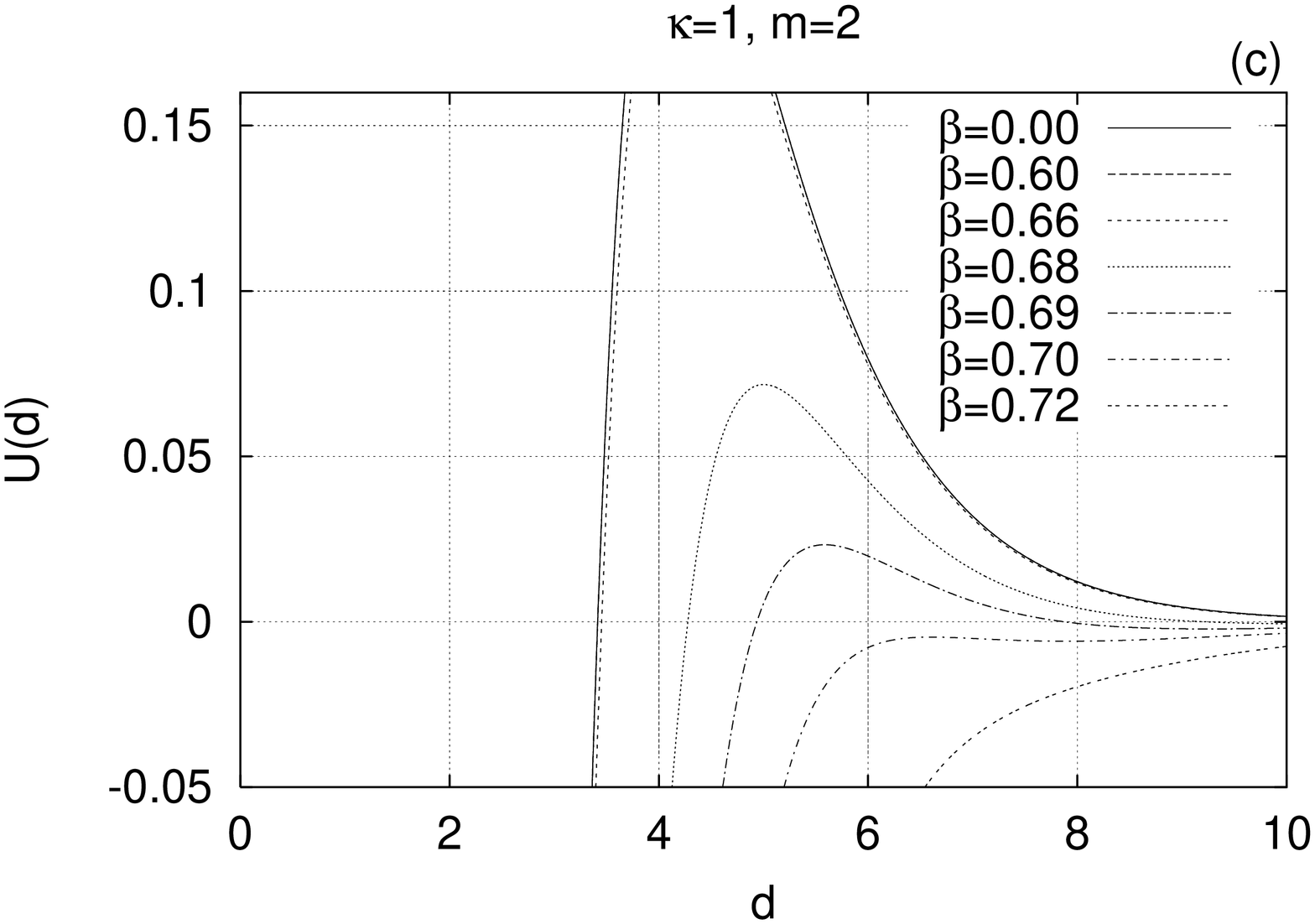}
\includegraphics[width=4cm,angle=-90]{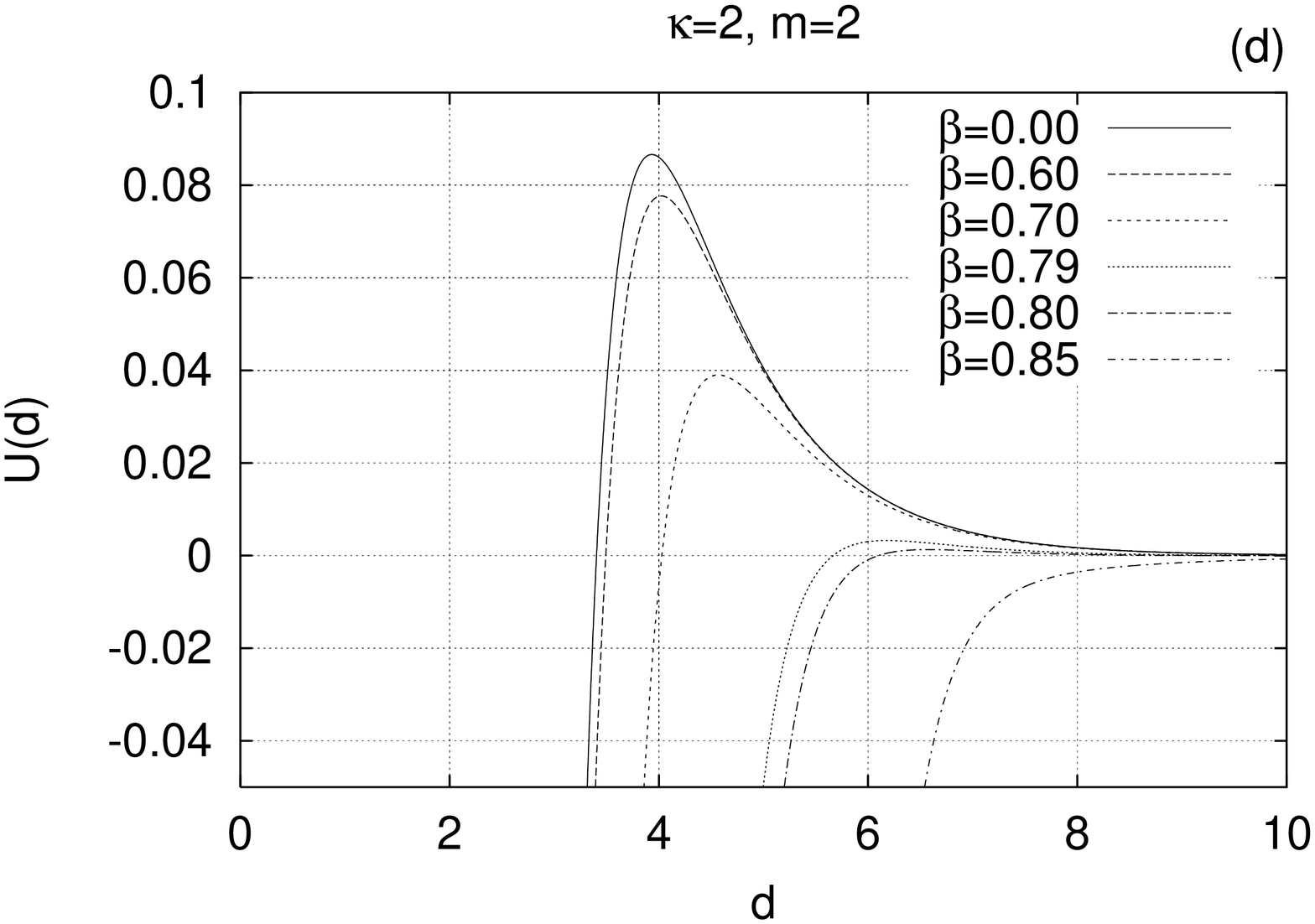}
\caption{\label{figso5}Intervortex
potential for $\kappa=1$ (a) and $\kappa=3$ (b) for $m=1$, 
and for $\kappa=1$ (c) and $\kappa=2$ (d) for $m=2$.}
\end{center}
\end{figure}

A couple of case studies will help illustrate the situation; it is
useful to see how the interaction energy varies with $\beta$ for fixed
$\kappa$ (see
Figure \ref{figso5}). In Figure \ref{figso5}b,
the case $\kappa=3.0$ 
is displayed. (This corresponds to the second of the three cases
mentioned near the beginning of this section.) The first two curves
($\beta=0.0$ and $\beta=0.8$) coincide, since the core has yet to
become AF (as can be ascertained from Figure \ref{kbplane}, since these points lie
to the left of Curve A). The subsequent curves show a decrease in the
potential as the $\eta$ field turns on, along with its attractive
contribution to the vortex energy. Eventually when
$\beta\simeq 0.92$, the potential goes from repulsive to
attractive, in agreement with (\ref{betastar}), and also with
(\ref{kc}).

In Figure \ref{figso5}a, 
the case $\kappa=1.0$ (the third case mentioned near the beginning of
this section) is displayed. The first two curves, corresponding to
parameters lying to the left of Curve A in Figure \ref{kbplane}, coincide, since
the core of the vortices is normal; subsequent increases in $\beta$
produce a drop in the interaction energy, as expected since $\eta$ now
contributes. It is important to note, however, that for $\kappa=1.0$ the
superconductor is type I starting at $\beta_c=0.33$. We would thus
expect the vortices to attract for $\beta>0.33$. This is not the case:
as can be seen from Figure \ref{figso5}a, the attraction begins only around
$\beta\simeq0.883$, somewhat to the right of Curve A ($\beta_{AF}=0.881$).
(That it is
somewhat to the right of Curve A is simply due to the fact that $\eta$
(at the center of the vortex, for example) evolves continuously with
$\beta$ and is strictly zero up until Curve A; it is therefore small
and will have a negligible effect on the vortex interaction until it
has time to grow to an appreciable value (see the values of $C_\eta$
given in Table 2).)
Thus, there is a
zone (between curves A and C) wherein the superconductor is type I,
but wherein nonetheless vortices repel one another: this is the
identity crisis referred to earlier. The point is that
in this region the $\eta$ field {\em would} produce an
attractive long-range interaction, but it has not yet attained a
nonzero value in the core and therefore
has no effect on the vortex interaction.

The situation is even stranger if we consider vortices of winding 
number 2, for which Curve B delineates normal (to the left) vs. AF
(to the right) cores. For $\kappa=2.0$ (Figure \ref{figso5}d),
the vortex turns AF at 
$\beta_{AF}=0.562$, but vortices remain repulsive until the value
corresponding to the type I-type II transition ($\beta_c\simeq0.78$).
For $\kappa=1.0$ (Figure \ref{figso5}c), the vortices develop an AF core at
roughly $\beta_{AF}=0.659$ but remain repulsive until just to the right of
Curve C, as expected given that Curve C separates type II from type I
superconductivity. However, at a lower value of $\kappa$ we would find
that the interaction energy of 2-vortices in Region 6 is repulsive,
which is incommensurate with the fact that the superconductor is type
I there.

Note that these ``identity crises'' only occur for certain values of
the parameters, and in particular for $\kappa<2.25$, so this does not 
appear to be relevant to HTSC (supposing that the SO(5) model
provides a good description of these systems),
since all known HTSCs have quite large
values of $\kappa$ ($\kappa\gtwid50$).
However, in the case of superconducting cosmic strings 
and optical vortex solitons, there are no restrictions on the value of 
$\kappa$ and the effect may be relevant.  

It is interesting to note that the results obtained above complement
those obtained in previous work where the conditions under which SO(5) 
superconductors are type I or type II were determined. In   
Ref.\cite{Juneau:2001vz}, the distinction between type I and type II 
behaviour was determined by analysing the energy per unit flux of vortices 
as a function of their winding number. In 
Ref.\cite{Juneau:2002ax}, this distinction
was obtained by studying the energy density of a surface separating 
superconducting and normal regions at critical applied magnetic fields. The 
results stemming from both methods are in agreement and reveal, unexpectedly,
that superconductors whose $\kappa$ values would normally be
associated with type II superconductivity may in fact be
type I superconductors. The surprising region is $\kappa>1/\sqrt2$
(conventionally type II) and to the right of Curve C in Figure \ref{kbplane}.
The underlying reason is the possibility of an 
AF vortex core, or equivalently, the fact that an applied magnetic field
induces a transition between SC and AF (not normal) phases.  

\section{Conclusion}

In this paper we have derived an expression for the potential energy 
of two widely separated vortices within the framework of a model with
two order parameters, one of which is complex and attains a nonzero
expectation value. Examples of physical contexts where such a
situation occurs are bosonic superconducting
cosmic strings, optical vortex solitons and
the SO(5) model of HTSC.
The approach used was a point vortex approximation, due to Speight
\cite{Speight:1997sh}. Our starting point was a linearization of the
free energy (\ref{hf}) to which source terms (\ref{hs}) were added;
the form of the sources was given in (\ref{cf},\ref{ca}).
The main result is the two-vortex interaction energy, (\ref{pot}). The
constants appearing in this expression were chosen so that the
long-range fields of a vortex in the point vortex approximation agree
with the full nonlinear model.

The behaviour of the vortices is for the most part in agreement with
previous work, described in Section 2, namely, that for fixed
$\kappa$, type II behaviour (stable vortex lattice) turns into type I
behaviour (unstable lattice) for $\beta$ sufficiently close to 1,
provided that the vortices have AF cores.

This caveat, although apparently innocent, is actually quite
important, and yields a surprising result. Namely, for certain regions
in parameter space (regions 5 and 6 in Figure \ref{kbplane}), the superconductor
is type I, yet vortices have normal (not AF) cores and repel one
another, a behaviour normally associated with type II
superconductors. It would appear that in such a case a vortex lattice
would be metastable rather than stable (the free energy of a
macroscopic non-SC region containing the same flux being lower than
that of the vortex lattice). It is ironic and amusing that the
surprising behaviour uncovered in \cite{Juneau:2001vz,Juneau:2002ax}
(type I SC for
$\kappa>1/\sqrt2$) can be thought of as being due to the fact that 
vortices have AF cores, whereas the surprising behaviour uncovered
here (vortices which repel one another in type I SC) is due to the
fact that vortices have normal, not AF cores.

While the attraction or repulsion of vortices is clearly of interest
in any physical situation described by the above work, the unexpected
behaviour uncovered here is unlikely to be seen in HTSC, as it occurs
in the wrong region of parameter space.

\vspace{.5cm}
This work was supported by the Natural Science and Engineering
Research Council of Canada and by the Fonds de Recherche sur la Nature
et les technologies du Qu\'ebec.

\bibliographystyle{apsrev}

\bibliography{fieldtheory}

\end{document}